\documentclass[12pt,twoside]{article}
\usepackage{fleqn,espcrc1}

\usepackage{epsfig}

\newcommand{\beq}{\begin{equation}}
\newcommand{\eeq}{\end{equation}}
\newcommand{\bea}{\begin{eqnarray}}
\newcommand{\eea}{\end{eqnarray}}

\newcommand{\noi}{\noindent}

\title{$J/\psi$ suppression at SPS and RHIC in the comovers approach}

\author{N. Armesto\address{Universidad de C\'ordoba, E-14071 C\'ordoba, Spain},
         A. Capella\address[MCSD]{LPT, Universit\'e de Paris XI, 
B\^atiment 210, 91405 Orsay Cedex,
France}\thanks{Unit\'e Mixte de Recherche UMR n$^{\circ}$ 8627 - CNRS},
         E. G. Ferreiro\addressmark[MCSD],
         A. Kaidalov\address{ITEP,
B. Cheremushkinskaya ul. 25, 117259 Moscow, Russia}
         and
          D. Sousa\addressmark[MCSD]}
 
\begin{document}
\maketitle

\noindent{\it (presented by A. Capella)}

\begin{abstract}
The NA50 collaboration data on the $J/\psi$ suppression are compared with the
results obtained in a comovers approach based on the Dual Parton Model (DPM).
Predictions for the $J/\psi$ suppression versus the charged multiplicity --
measured in the rapidity region of the dimuon trigger -- are given for SPS and
RHIC energies.
  \end{abstract}
\vskip 0.1 truecm

\setcounter{footnote}{0}

The NA50 collaboration has observed an anomalous $J/\psi$ suppression in
$Pb$ $Pb$ collisions i.e. a suppression that exceeds the one expected from
the extrapolation of the $pA$ and $SA$ data, and exhibits an interesting
centrality pattern. These data \cite{1r} have been interpreted either as the
result of one (or two) deconfining phase transition(s) or as due to the
interaction with comovers. Our aim is to examine to what extent the observed
pattern can be reproduced in the latter approach. Here, the $J/\psi$ survival
probability is the product of two factors $S_{abs}(b,s)\cdot S_{co}(b,s)$. The
first factor represents the suppression due to nuclear absorption of the
$c\bar{c}$ pair. Its expression, given by the probabilistic Glauber model, is
well known. It contains a parameter, the absorptive cross-section
$\sigma_{abs}$. The second factor $S_{co}(b,s)$ represents the suppression
resulting from the interaction with comovers. Its expression \cite{2r,3r,4r} is

\beq
\label{1e}
S_{co}(b,s) = \exp \left \{ - \sigma_{co} \ N_{y_{DT}}(b, s) \ \ln { \left [
{N^{co}_{y_{DT}}(b, s) \over N_{f}} \right ] } \right \} \ .
\eeq

\noi Here $N_y^{co}$ represents the density of comovers (positive negative and
neutrals) in the rapidity region of the dimuon trigger ($0 < y^* < 1$). We
determine it in the Dual Parton Model (see Eq. (\ref{3e}) below). $N_f =
1.15$~fm$^{-2}$ is the corresponding density in $pp$, and $\sigma_{co}$ is an
averaged interaction cross-section. In the calculations below we take
$\sigma_{co} = 1$~mb and $\sigma_{abs} = 4.5$~mb. \par

In this way we can compute the $J/\psi$ suppression at each impact parameter.
However, experimentally, the ratio of $J/\psi$ over DY is plotted as a function
of either $E_T$ or the energy of the zero degree calorimeter $E_{ZDC}$. $E_T$
is the transverse energy of neutrals deposited in the NA50 calorimeter, located
in the backward hemisphere ($1.1 < y_{lab} < 2.3$). Using the proportionality
between $E_T$ and multiplicity, we have

\beq
\label{2e}
E_T(b) = {1 \over 3} \ q \ N_{y_{cal}}^{co}(b) \quad .
\eeq

  \noi Here the multiplicity of comovers is
determined in the rapidity region of the NA50 calorimeter. The factor 1/3 is
introduced because only the energy of neutrals is recorded. In this way $q$ is
close to the transverse energy per particle, but it contains also the 
calibration
of the calorimeter. From a fit to the $E_T$
distribution we obtain $q = 0.6$~GeV. An alternative determination of $q$ is
obtained from the measured correlation between $E_T$ and $E_{ZDC}$ (see below).
The latter is defined as $E_{ZDC}(b) = [A - n_A(b)] E_{in}$. Here $n_A$ is the
number of participants of nucleus, $A - n_A$ the number of spectators, and
$E_{in}$ = 158~GeV is the beam energy. We see from Eqs. (\ref{1e}) and
(\ref{2e}) that, in order to describe the centrality dependence of the $J/\psi$
suppression, it is paramount to have a good description of the $b$ dependence
of $N_y^{co}$ -- both in the rapidity region of the dimuon trigger and in the
one of the $E_T$ calorimeter. The $b$-dependence of charged multiplicity at
mid-rapidities has been measured at SPS and RHIC. It has been shown in
\cite{5r} that these data are well described in DPM, where the multiplicity is
given by a linear combination of the number of participants and the number of
binary collisions

\beq
\label{3e}
N_y^{co}(b) = A_y(b) \ n_A(b) + B_y(b) \ n(b)
\eeq

\noi with coefficients that depend on $b$ and $y$ and can be
calculated in the model. Their values at mid-rapidities are given in
\cite{5r}. Using Eq. (\ref{3e}), with coefficients calculated in the rapidity
region of the calorimeter, we obtain \cite{6r} a good description of 
the measured $E_T -
E_{ZDC}$ correlation with $q = 0.6$~GeV (Fig.~1)\footnote{In \protect{\cite{2r,3r,4r}}
an approximation was made in which the dependence on $b$ of the 
coefficients $A$
and $B$ was dropped. Moreover, a two-string approximation was used for each
$NN$-collision. It turns out that these approximations have a very small effect
on the ratio $J/\psi$ over DY at each $b$. However, they do change the
correlation between $b$ and $E_T$ -- or between $E_T$ and $E_{ZDC}$. As a
consequence, an extra term in the r.h.s. of Eq. (\ref{2e}) was needed in
\protect{\cite{3r,4r}} in order to reproduce this correlation.}. The
correlation $E_T - b$ is given by $P(E_T , b) \propto \exp \{ - (E_T -
E_T(b))^2/2qaE_T(b)\}$ with $q = 0.6$~GeV and $a = 0.88$. \par

The model allows to compute the ratio $J/\psi$ over DY versus either $E_T$
or $E_{ZDC}$ from peripheral collisions up to the knee of the $E_T$
distribution. To go beyond it, we have to introduce \cite{4r,7r} the
fluctuations responsible for the tail of the $E_T$ distribution (Eq.
(\ref{2e}) gives only the average value of $E_T$ at each $b$). They have
been introduced \cite{4r} in the model by multiplying
$N_{y_{DT}}^{co}(b)$ in Eq. (\ref{2e}) by $F(b) = E_T/E_T(b)$, where
$E_T$ is the measured value of the transverse energy. \par

The results are presented in Fig.~2a and compared with the NA50 data. We
see that from peripheral collisions up to the knee of the $E_T$
distribution, the data are well described. However, beyond
the inflexion point at the knee, the decrease in the data is sharper
than in the model. Note, however, that the data beyond the knee are
obtained with the so-called minimum bias (MB) analysis. Only the ratio
$J/\psi$ over MB is measured and it is multiplied by a theoretical
ratio DY/MB. In the model, this ratio (as well as the ratio
$J/\psi$/MB) is essentially flat beyond the knee -- due to the fact
that the tail of the $E_T$ distribution of hard ($J/\psi$, DY) and
soft (MB) processes is assumed to be the same. In sharp contrast with
this behaviour, the measured ratio $J/\psi$ over MB has a very
pronounced turn-over at the knee. It is most important to determine
whether such a turn-over is due to an increase in the $J/\psi$
suppression, or, on the contrary, to a small difference in the widths
of the tails of the $E_T$ distributions of soft and hard processes.
In the latter case a similar turn-over would also be present in the
ratio DY/MB -- resulting in a flatter behaviour of the ratio
$J/\psi$ over DY beyond the knee. On the theoretical side the
calculation beyond the knee is also subject to uncertainties. In
particular, the rapidity regions of the dimuon trigger and the $E_T$
calorimeter are far apart and the fluctuations in the two regions
could be different. In view of that, it would be very interesting to
measure the $J/\psi$ suppression versus charged multiplicity
-- with the dimuon and multiplicity triggers sitting in the same
rapidity region. The results of the model \cite{6r} 
for SPS and RHIC are shown
in Fig.~3. \par

The NA50 collaboration has presented new data on the $J/\psi$
suppression versus $E_{ZDC}$. In so far as a theoretical model
describes well the $E_T - E_{ZDC}$ correlation, a set of data can be
plot versus either variable without further constraints on the
model. However, the new data are important since they include many
peripheral points. The results of the model are presented in Fig.~2b. The
centrality dependence of the $J/\psi$ suppression is reasonably
well reproduced. However, the absolute normalization used in Fig.~2a 
is here 10 \% too large.
(Note that the normalization of the data in Fig.~2b is not measured. 
It has been fixed
\protect{\cite{1r}} from the one in Fig.~2a using only points in the 
range $60 < E_T < 100$~GeV).
This point has important consequences for the interpretation of the 
data. Indeed, it is obvious
from a comparison of figures 2a and 2b that, if the relative 
normalization of the two sets of data
were correct, the onset of the anomalous suppression in $E_{ZDC}$ 
would take place at a value of
$b$ significantly larger than in $E_T$. This problem is solved with 
the change of normalization
discussed above. However, in this case there would be several 
experimental points
sitting above the NA50 nuclear absorption curve. More
important, independently of 
their absolute normalization,
the data of Fig.~2b for $E_{ZDC} < 28$~TeV exhibit a $J/\psi$ 
suppression significantly steeper
than the NA50 absorption model -- indicating that the anomalous 
suppression is already present
in very peripheral collisions, where the density is significantly
lower than the maximal one in S-U.

In conclusion, the NA50 data on the $J/\psi$ suppression from peripheral
collisions up to the knee of the $E_T$ distribution can be described in a
comovers approach. (In contrast, both the NA50 absorption model and 
the model \cite{7r}
fail to reproduce the peripheral data). Beyond the knee the obtained 
suppression is too small.
However, in this region the theoretical uncertainties are large and 
the data contain a
theoretical input that has to be checked experimentally.
 
\vskip 0.4 truecm

This work was supported by NATO grant PSTCLG 977275.

\begin{figure}[t]
\begin{minipage}[t]{80mm}
\center\epsfig{file=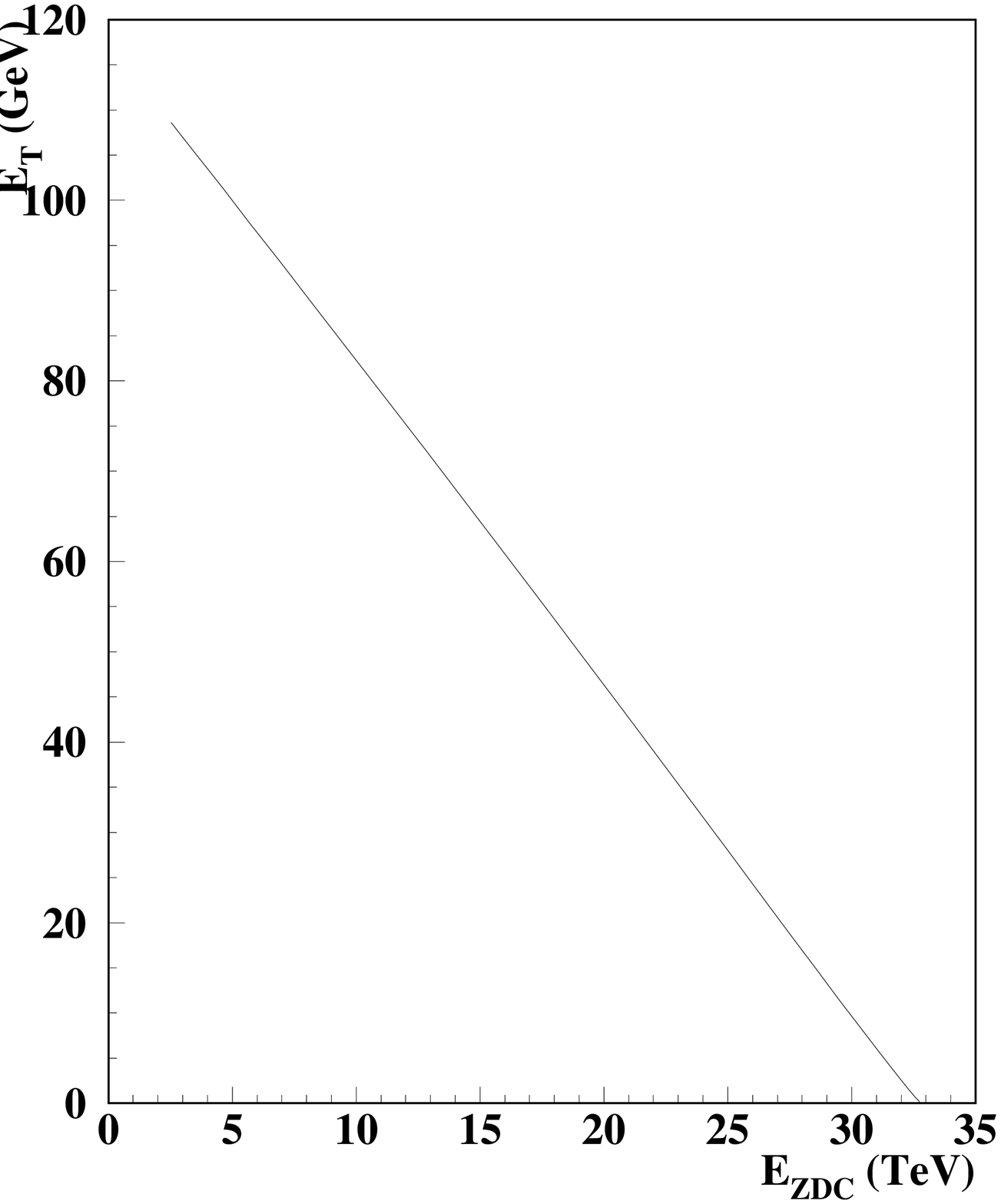,bbllx=0,bblly=0,bburx=540,bbury=560,height=5.3cm}
\caption{$E_{T}-E_{ZDC}$ correlation.}
\end{minipage}
\hspace{\fill}
\setcounter{figure}{2}
\begin{minipage}[t]{75mm}
\center\epsfig{file=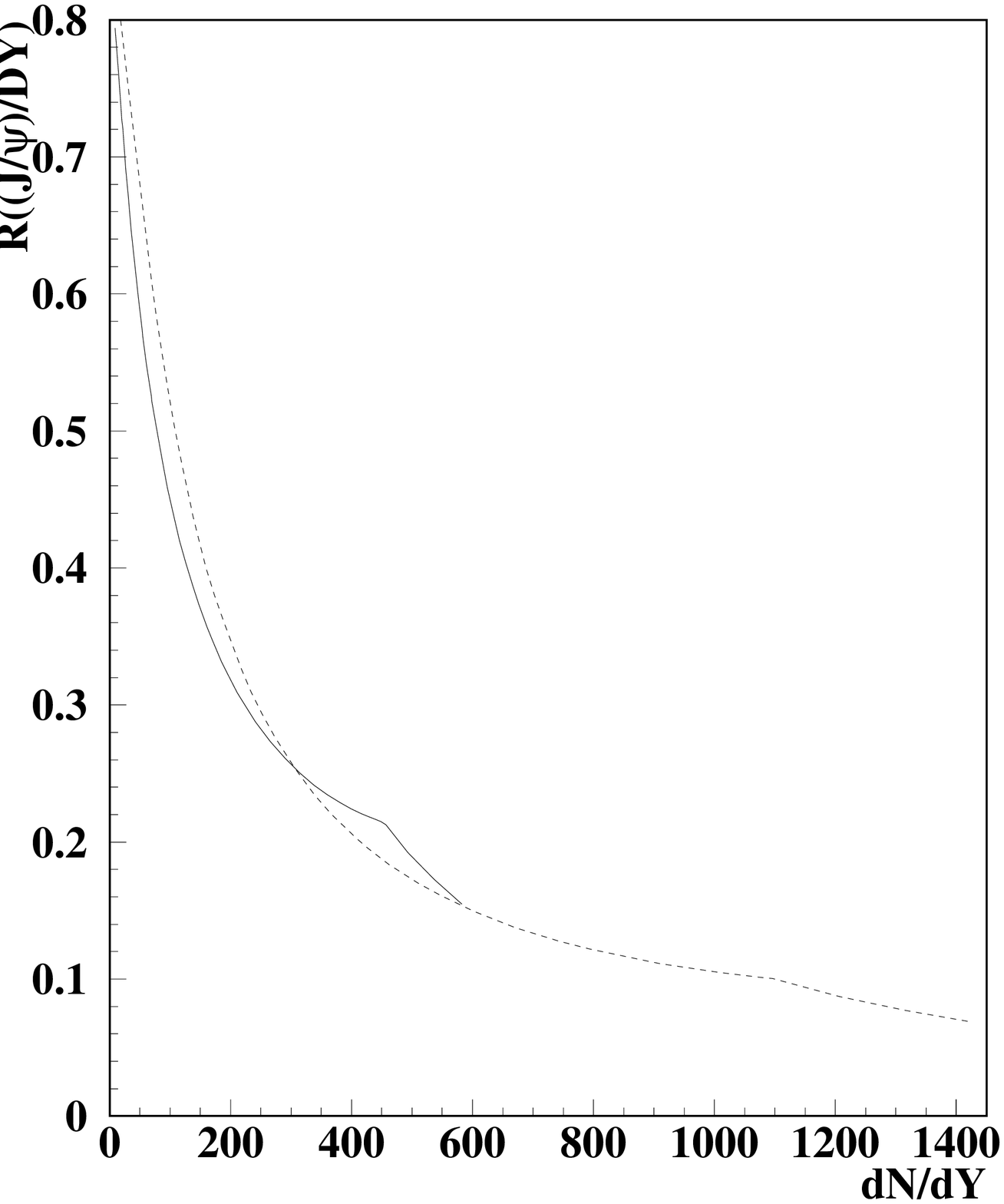,bbllx=0,bblly=0,bburx=540,bbury=560,height=5.3cm}
\caption{Ratio $J/\psi$ over DY 
normalized to $pp$
versus charged multiplicity in Pb-Pb collisions at $p_{lab} =$ 158
AGeV/c, both in the range $0 < y^* < 1$, (full line) and 
at $\sqrt{s} =$ 200 AGeV in the range $- 0.5 < y^* < 0.5$ (dotted line).}
\end{minipage}
\end{figure}

\setcounter{figure}{1}

\begin{figure}[b]
\begin{minipage}[t]{80mm}
\centering\epsfig{file=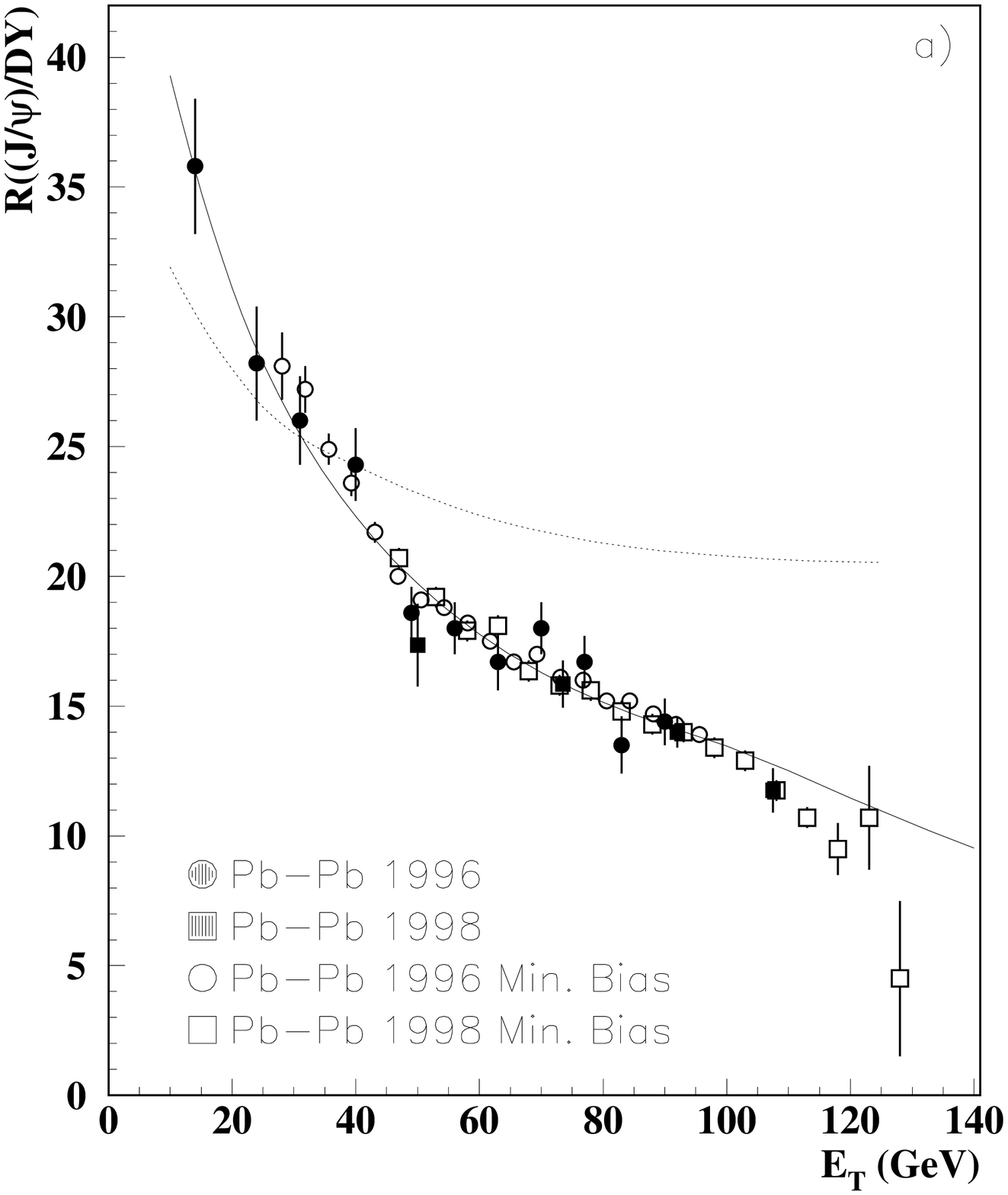,bbllx=0,bblly=0,bburx=540,bbury=560,height=7.2cm}
\end{minipage}
\hspace{\fill}
\begin{minipage}[b]{80mm}
\centering\epsfig{file=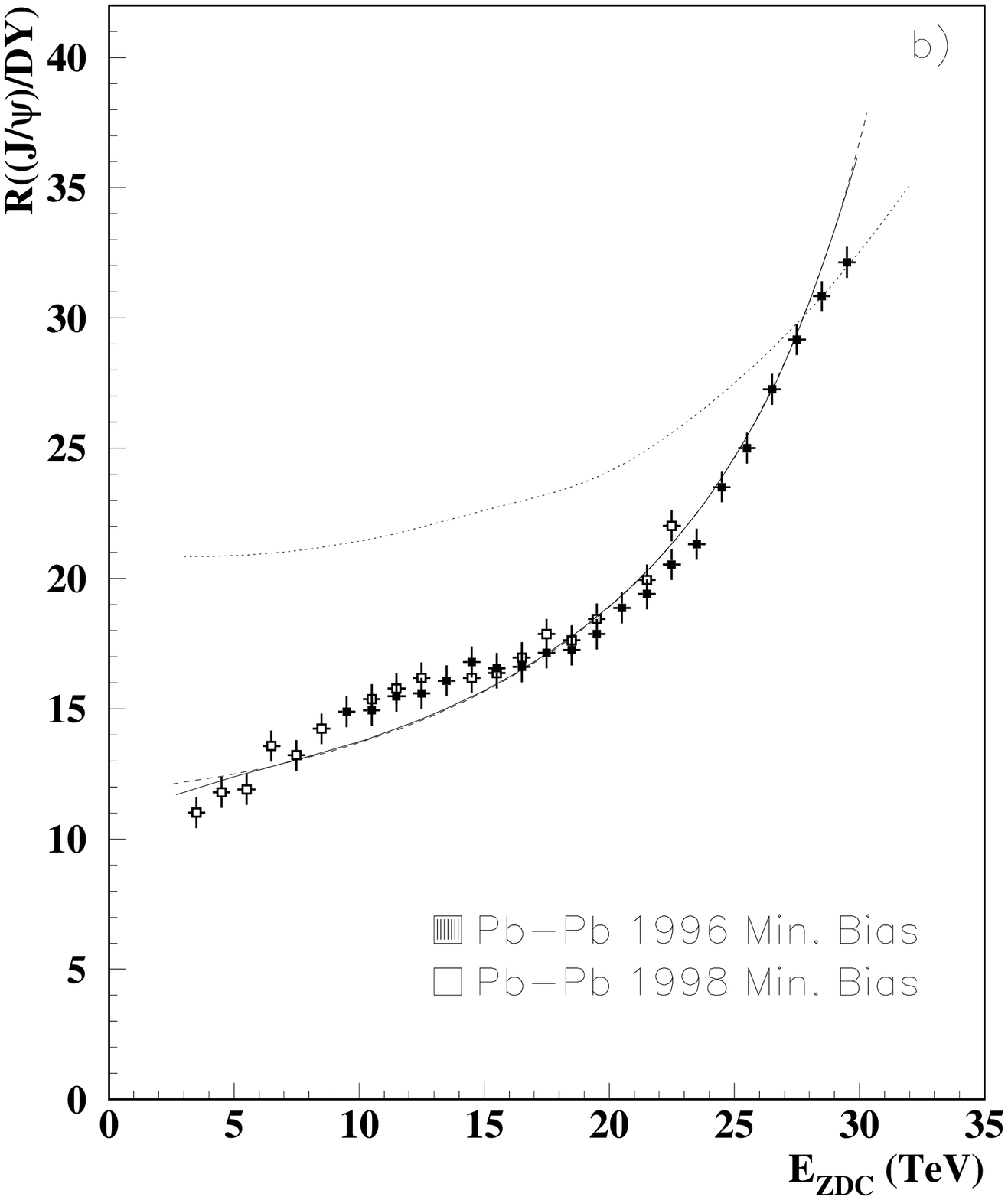,bbllx=0,bblly=0,bburx=540,bbury=560,height=7.2cm}
\end{minipage}
\caption{a): Ratio $J/\psi$ over DY versus $E_T$ compared to NA50 data
\protect{\cite{1r}}. The full curve is the theoretical prediction.
b): Ratio $J/\psi$ over DY versus $E_{ZDC}$ compared to preliminary
results presented by NA50 \protect{\cite{1r}}. The full line is obtained from
the full line in a) using the calculated $E_T - E_{ZDC}$ correlation and changing
the normalization by a factor 0.92 (see text). The dashed line is obtained computing for
each $b$, the value of $E_{ZDC}$ and the value of the ratio $R$ -- with the same
change in normalization. In both figures the dotted line is the NA50 absorption model,
fitting pA and SU.} 
\end{figure}


\begin{thebibliography}{9}
\bibitem{1r} NA50 collaboration, M. C. Abreu et al, Phys. Lett. B477 (2000) 28;
{\it ibid.} these Proceedings (presented by P. Bordalo).
\bibitem{2r} N. Armesto and A. Capella, Phys. Lett. B430 (1998) 23.
\bibitem{3r} N. Armesto, A. Capella and E. G. Ferreiro, Phys. Rev. C59 (1999)
359.
\bibitem{4r} A. Capella, E. G. Ferreiro and A. Kaidalov, Phys. Rev. Lett. 85
(2000) 2080.
\bibitem{5r} A. Capella and D. Sousa, Orsay preprint LPT 00-137, 
nucl-th/0101023.
\bibitem{6r} A. Capella and D. Sousa, in preparation.
\bibitem{7r} J.-P. Blaizot, P. M. Dinh and J. Y. Ollitrault, Phys. Rev. Lett.
85 (2000) 4020.
\end{thebibliography}
\end{document}